\documentclass[aps,prd,preprintnumbers,showpacs,showkeys,nofootinbib,
superscriptaddress,fleqn,floatfix,tightenlines,10pt]{revtex4-1}
\usepackage{amsmath,amsfonts,amssymb,amscd,amsxtra,amsthm}
\usepackage{graphicx}  % Include figure files
\usepackage{epstopdf}
\usepackage{dcolumn}  % Align table columns on decimal point
\usepackage{bm}          % bold math
\usepackage{slashed}
\usepackage{cancel}
\usepackage{float}
\usepackage{mathtools}
\usepackage{amsbsy}
\usepackage{amstext}

\usepackage[utf8]{inputenc} 
\usepackage{booktabs}
\usepackage[normalem]{ulem} % \sout{old text} for strikeout
\usepackage[dvipsnames]{xcolor} % For blue in-text comments and
\usepackage{tabularx}
\usepackage{enumitem}  
\usepackage{array} 
\usepackage{slashed}
\usepackage{tikz}
\usepackage{float}
\usepackage{multirow}
\renewcommand\sout{\bgroup \color{red} \ULdepth=-.5ex \ULset}

\makeatletter

\begin{document}
%===================================================
\preprint{INHA-NTG-15/2018}
\title{$\eta n$ photoproduction and nucleon resonances} 
\author{Jung-Min Suh}
\email[E-mail: ]{suhjungmin@inha.edu}
\affiliation{Department of Physics, Inha University, Incheon 22212,
 Republic of Korea}
%-------------------------------------------------------------------------
\author{Sang-Ho Kim}
\email[E-mail: ]{sangho\_kim@korea.ac.kr}
\affiliation{Center for Extreme Nuclear Matters (CENuM),
Korea University, Seoul 02841, Republic of Korea}
\affiliation{Department of Physics, Pukyong National University
  (PKNU), Busan 48513, Republic of Korea}
%-------------------------------------------------------------------------
\author{Hyun-Chul Kim}
\email[E-mail: ]{hchkim@inha.ac.kr}
\affiliation{Department of Physics, Inha University, Incheon 22212,
 Republic of Korea}
\affiliation{Advanced Science Research Center, Japan Atomic Energy
  Agency, Shirakata, Tokai, Ibaraki, 319-1195, Japan} 
\affiliation{School of Physics, Korea Institute for Advanced Study 
 (KIAS), Seoul 02455, Republic of Korea}
%------------------------------------------------------------------------- 
\date{\today}

\begin{abstract}
  We present results of a recent work on the reaction mechanism of
  $\eta n$ photoproduction off the neutron in the range of $ \sqrt{s}
  \approx 1.5 - 1.9$ GeV and discuss the role of various nucleon
  resonances listed in the Particle Data Group (PDG). 
  We make use of an effective Lagrangian approach combining with a Regge
  method. The total and helicity-dependent cross sections
  $\sigma_{1/2},\,\sigma_{3/2}$ are computed and the numerical results
  are in good agreement with the A2 experimental data. 
% We would like to encourage you to list your keywords within
% the abstract section using the \keywords{...} command.
\end{abstract}
\maketitle

\section{Introduction}
$\eta n$ photoproduction is one of the most practical and useful reaction
processes to investigate various $N^*$ resonances.
More interesting is that the narrow bump-like structure near $\sqrt{s} \sim
1.68$ GeV is seen only at this neutron channel, which was coined as
the neutron 
anomaly~\cite{Kuznetsov:2006kt,Jaegle:2008ux,Witthauer:2017pcy,Werthmuller:2013rba,Witthauer:2013tkm,Werthmuller:2014thb,Witthauer:2017get,Witthauer:2017wdb}. 
A number of theoretical studies have been performed whether this
phenomena is due to the existence of the narrow resonance
$N(1685,1/2^+)$ or not. In this talk, we revisit the $\gamma n \to
\eta n$ process, taking account 
of a total of fifteen $N^*$ resonances given in the PDG and the
$N(1685,1/2^+)$ additionally.
All the resonance parameters were extracted from the experimental data
on the PDG~\cite{Tanabashi:2018} and quark model
predictions~\cite{Capstick:1998uh}. 
The numerical results show that the $N(1685,1/2^+)$ is an
essential part of describing the $\eta n$ photoproduction together with
the dominant $N(1535,1/2^-)$ resonance.

\section{Formalism}
\begin{figure}[htp]
\centering
\includegraphics[width=9cm]{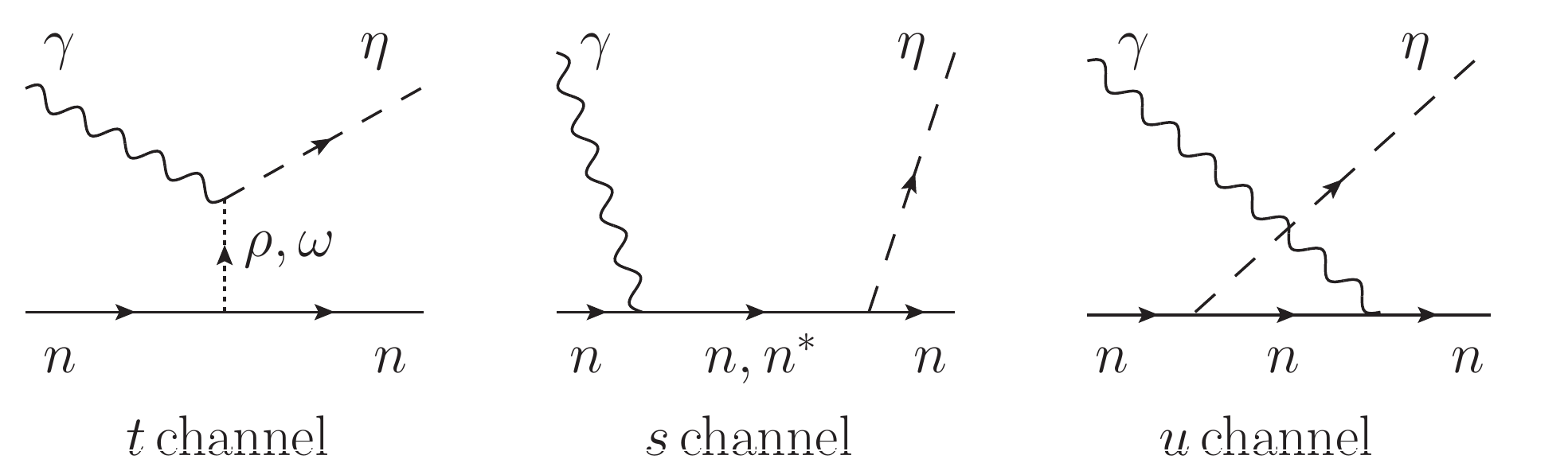}
\caption{Feynman diagrams for the $\gamma n \to \eta n$ reaction.}
\label{fig:1}
\end{figure}

The Feynman diagrams for the $\gamma n \to \eta n$ reaction are drawn
in Fig.~\ref{fig:1}. As a background contribution, we take into
account $\rho$ and $\omega$ Reggeon exchanges in the $t$ channel, and 
$N$ exchanges in both the $s$ and $u$ channels. 
We can construct the effective Lagrangians for the photon
(meson) and nucleon vertices as follows 
\begin{eqnarray}
\mathcal L_{\gamma NN} &=&
- \bar N \left[ e_N \gamma_\mu - \frac{e\kappa_N}{2M_N}
                           \sigma_{\mu\nu}\partial^\nu \right] A^\mu N,   \cr
\mathcal L_{VNN} &=&
-g_{VNN} \bar N \left[ \gamma_\mu N -
\frac{\kappa_{VNN}}{2M_N} \sigma_{\mu\nu} N
\partial^\nu \right] V^\mu + \mathrm{h.c.},                              \cr
\mathcal L_{\eta N N} &=&
\frac{g_{\eta N N}}{2 M_N} \bar N \gamma_\mu \gamma_5 N  \partial^\mu
                          \phi_\eta, 
\end{eqnarray}
where the values of coupling constants are fixed to be
$\kappa_N = -1.91$~\cite{Tanabashi:2018}, $g_{\rho NN} = 2.6$,
$\kappa_{\rho NN} = 3.7$, $g_{\omega NN} = 10.4$, $\kappa_{\omega NN}
= 0.41$, $g_{\eta NN} = 6.34$~\cite{Stoks:1999bz,Hohler:1974ht},
respectively. 

In addition to these Born terms, we include the fifteen $N^*$
resonances~\cite{Tanabashi:2018} and the narrow $N(1685,1/2^+)$ in the
$s$-channel diagram. We can calculate the corresponding transition
amplitudes from the following the effective Lagrangians 
\begin{eqnarray}
\mathcal L^{1/2^\pm}_{\gamma N N^*} &=& 
\frac{eh_1}{2M_N} \bar N \Gamma^{(\mp)}
\sigma_{\mu\nu} \partial^\nu A^\mu N^* + \mathrm{h.c.} ,               
\cr
\mathcal L^{3/2^\pm}_{\gamma N N^*}&=& 
-ie \left[ \frac{h_1}{2M_N} \bar N \Gamma_\nu^{(\pm)}
 - \frac{ih_2}{(2M_N)^2} \partial_\nu \bar N
 \Gamma^{(\pm)} \right] F^{\mu\nu} N^*_\mu + \mathrm{h.c.},   
\cr
\mathcal L^{5/2^\pm}_{\gamma N N^*} &=&
e\left[ \frac{h_{1}}{(2M_N)^2} \bar N \Gamma_\nu^{(\mp)}
-\frac{ih_{2}}{(2M_N)^3} \partial_\nu \bar N
\Gamma^{(\mp)} \right] \partial^\alpha F^{\mu\nu}
N^*_{\mu\alpha} + \mathrm{h.c.} ,
\cr
\mathcal L^{7/2^\pm}_{\gamma N N^*} &=&
ie \left[ \frac{h_{1}}{(2M_N)^3} \bar N \Gamma_\nu^{(\pm)}
-\frac{ih_{2}}{(2M_N)^4} \partial_\nu \bar N
\Gamma^{(\pm)} \right] \partial^\alpha \partial^\beta F^{\mu\nu}
N^*_{\mu\alpha\beta} + \mathrm{h.c.} ,
\end{eqnarray}
for the electromagnetic interactions and 
\begin{eqnarray}
\mathcal L^{1/2^\pm}_{\eta N N^*} &=&
 - i g_{\eta N N^*}\eta \bar N 
 \Gamma^{(\pm)} N^* + \mathrm{h.c.},                                   
\cr
\mathcal L^{3/2^\pm}_{\eta N N^*} &=&
 \frac{g_{\eta N N^*}}{M_\eta} \partial^\mu \eta \bar N
 \Gamma^{(\mp)} N^*_\mu + \mathrm{h.c.},
\cr
\mathcal L^{5/2^\pm}_{\eta N N^*} &=&
 \frac{ig_{\eta N N^*}}{M_\eta^2} \partial^\mu \partial^\nu \eta
 \bar N \Gamma^{(\pm)} N^*_{\mu\nu} + \mathrm{h.c.},
\cr
\mathcal L^{7/2^\pm}_{\eta N N^*} &=&
- \frac{g_{\eta N N^*}}{M_\eta^3} \partial^\mu \partial^\nu \partial^\alpha
\eta  \bar N \Gamma^{(\mp)} N^*_{\mu\nu\alpha} + \mathrm{h.c.} ,
\label{eq:ResLag2}
\end{eqnarray}
for the strong interactions with the notations
\begin{eqnarray}
 \Gamma^{{(\pm)}} = \left(
\begin{array}{c}
\gamma_5 \\ I_{4\times4}
\end{array} \right) ,
\,\,\,\,
\Gamma_\nu^{{(\pm)}} = \left(
\begin{array}{c}
\gamma_\nu \gamma_5 \\ \gamma_\nu
\end{array} \right) .
\label{eq:GammaPM}
\end{eqnarray}
Note that $N^*_\mu$, $N^*_{\mu\alpha}$, and $N^*_{\mu\alpha\beta}$
stand for the Rarita-Schwinger fields of spin-3/2, -5/2, and -7/2,
respectively. We refer to Ref.~\cite{Suh:2018yiu} for more details of
how the resonance couplings are determined.

\section{Results and discussion}
\begin{figure}[htp]
\centering
\includegraphics[width=5.7cm]{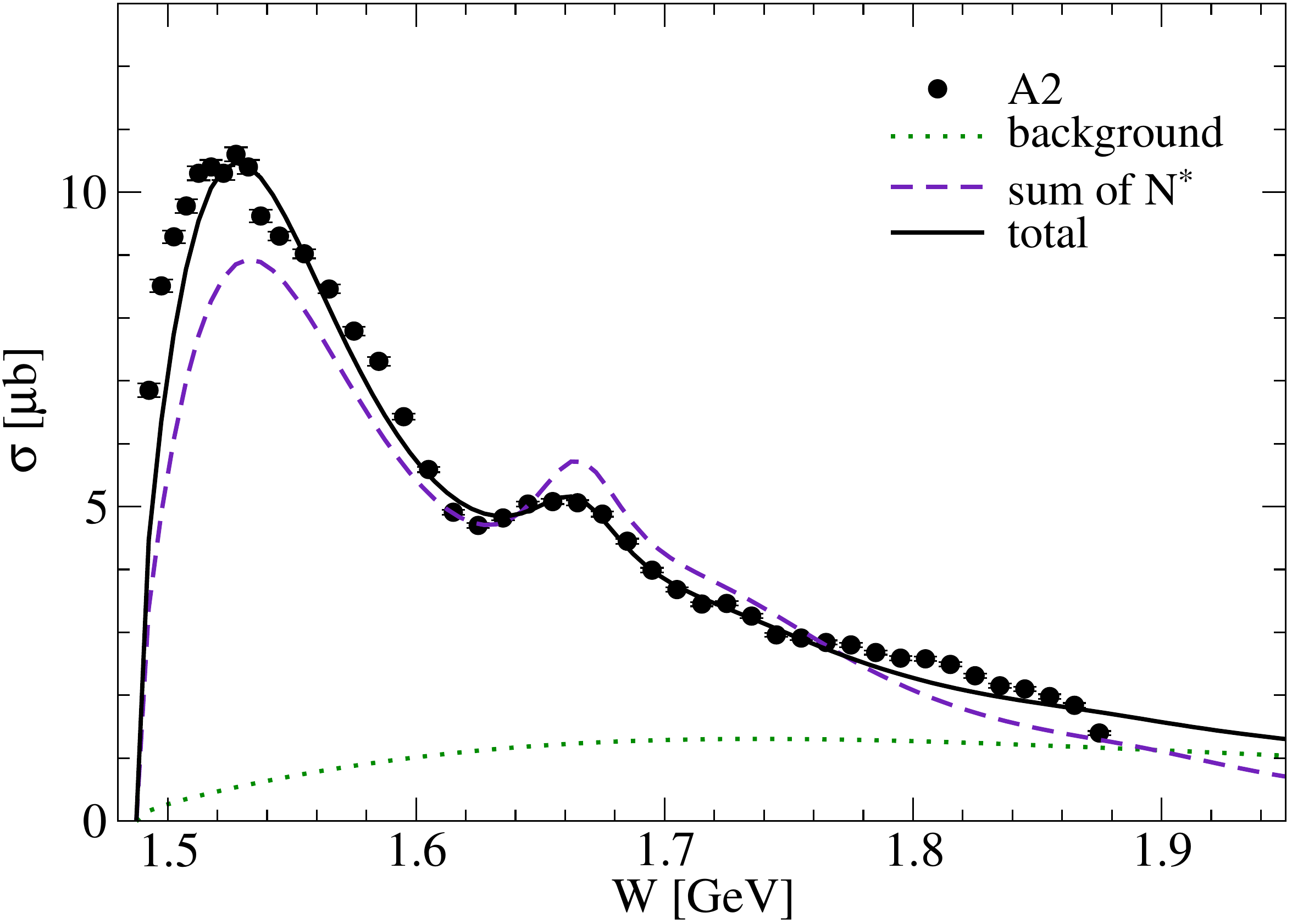}
\includegraphics[width=5.7cm]{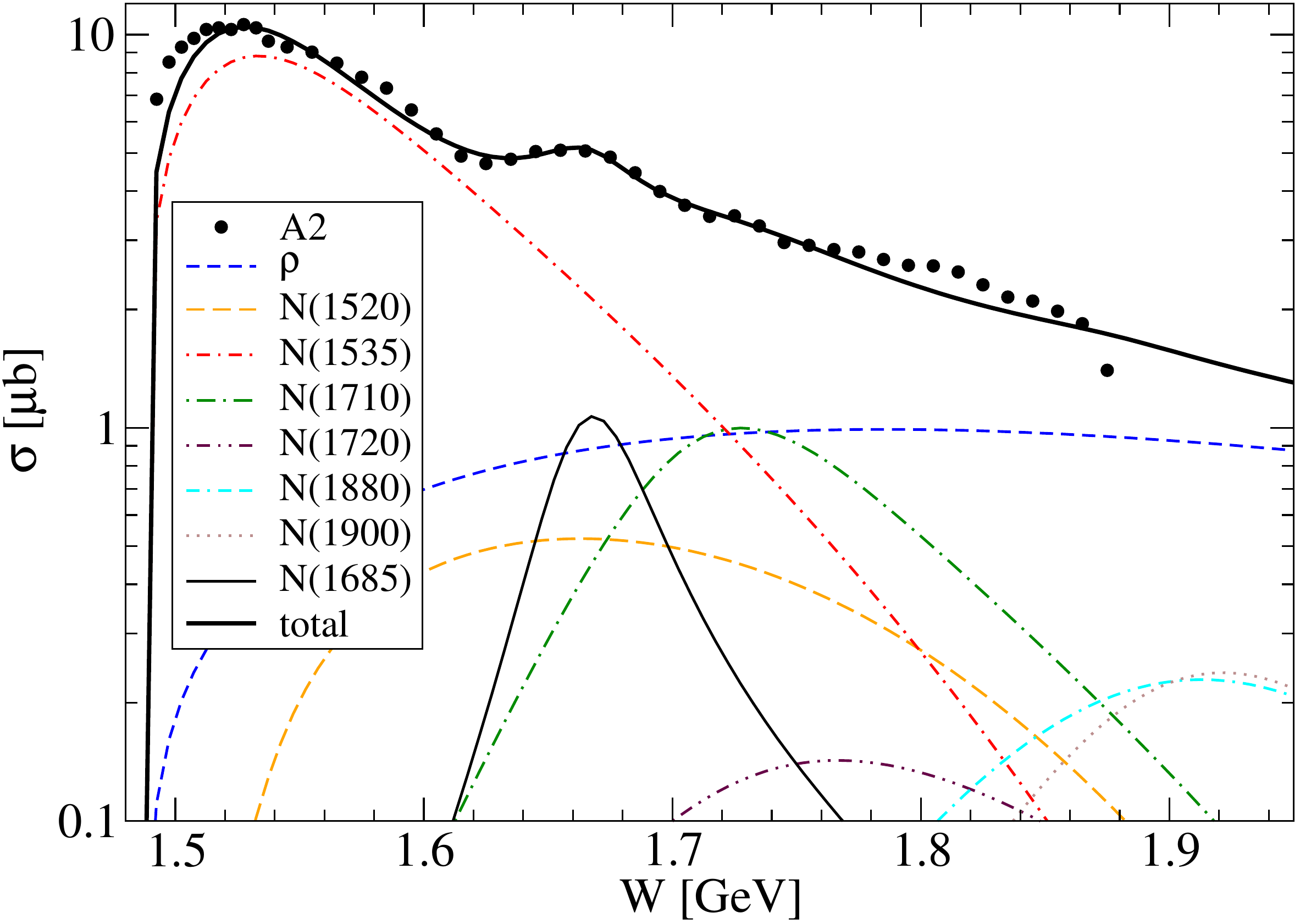}
\caption{Total cross section for $\gamma n \to \eta n$.
  The circles denote the A2 data~\cite{Werthmuller:2014thb}.}
\label{fig:2}
\end{figure}

Figure~\ref{fig:2} shows the numerical results of the total cross
section with various contributions. The total result describes the A2
data well~\cite{Werthmuller:2014thb}. The contribution of $\rho$
Reggeon exchange becomes large as $W$ increases. 
Meanwhile, the four $N^*$ resonances, i.e., $N(1520,3/2^-)$,
$N(1535,1/2^-)$, $N(1685,1/2^+)$, and $N(1710,1/2^+)$ are dominant
near threshold region. 

\begin{figure}[hbp]
\centering
\includegraphics[width=5.7cm]{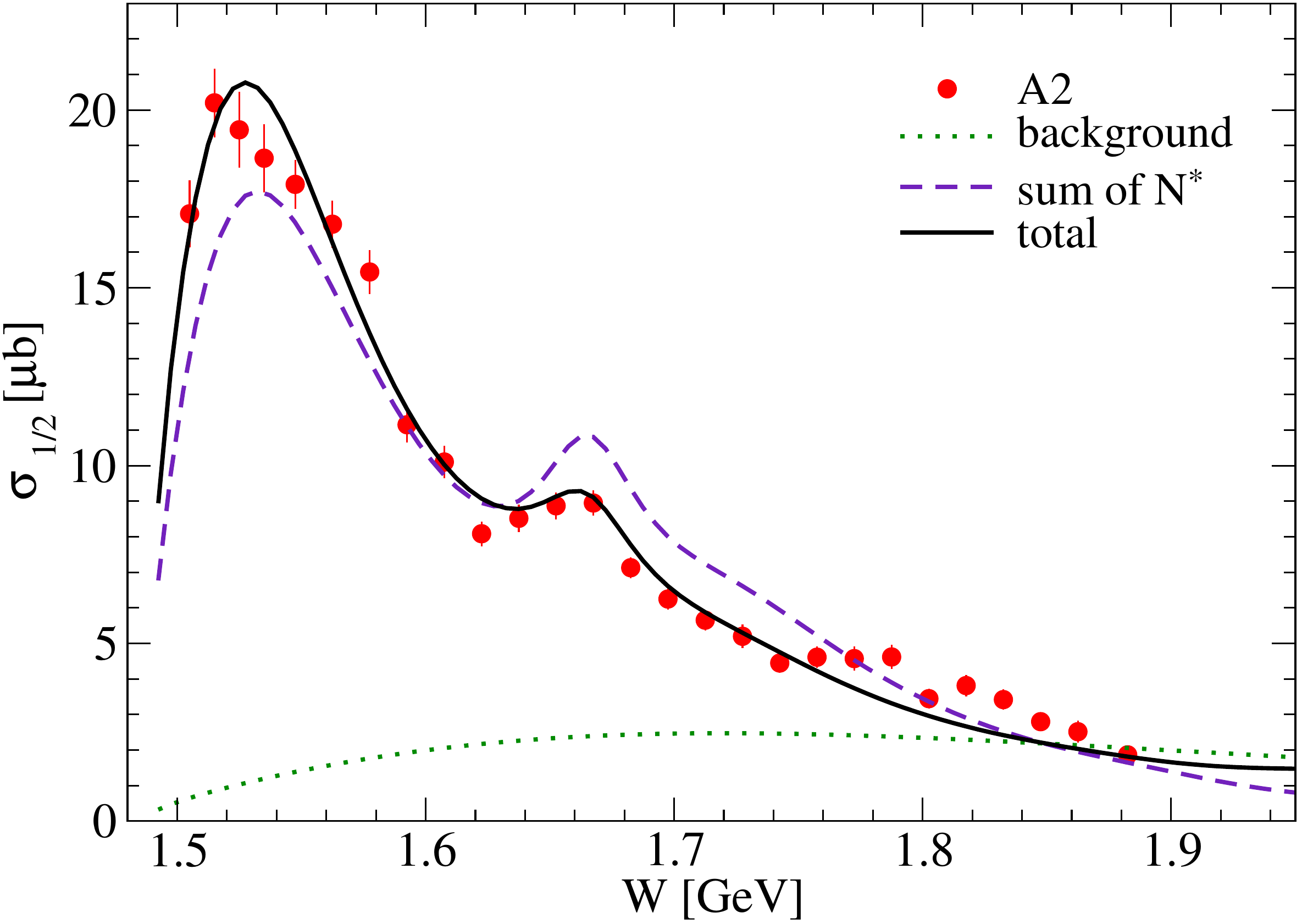}
\includegraphics[width=5.7cm]{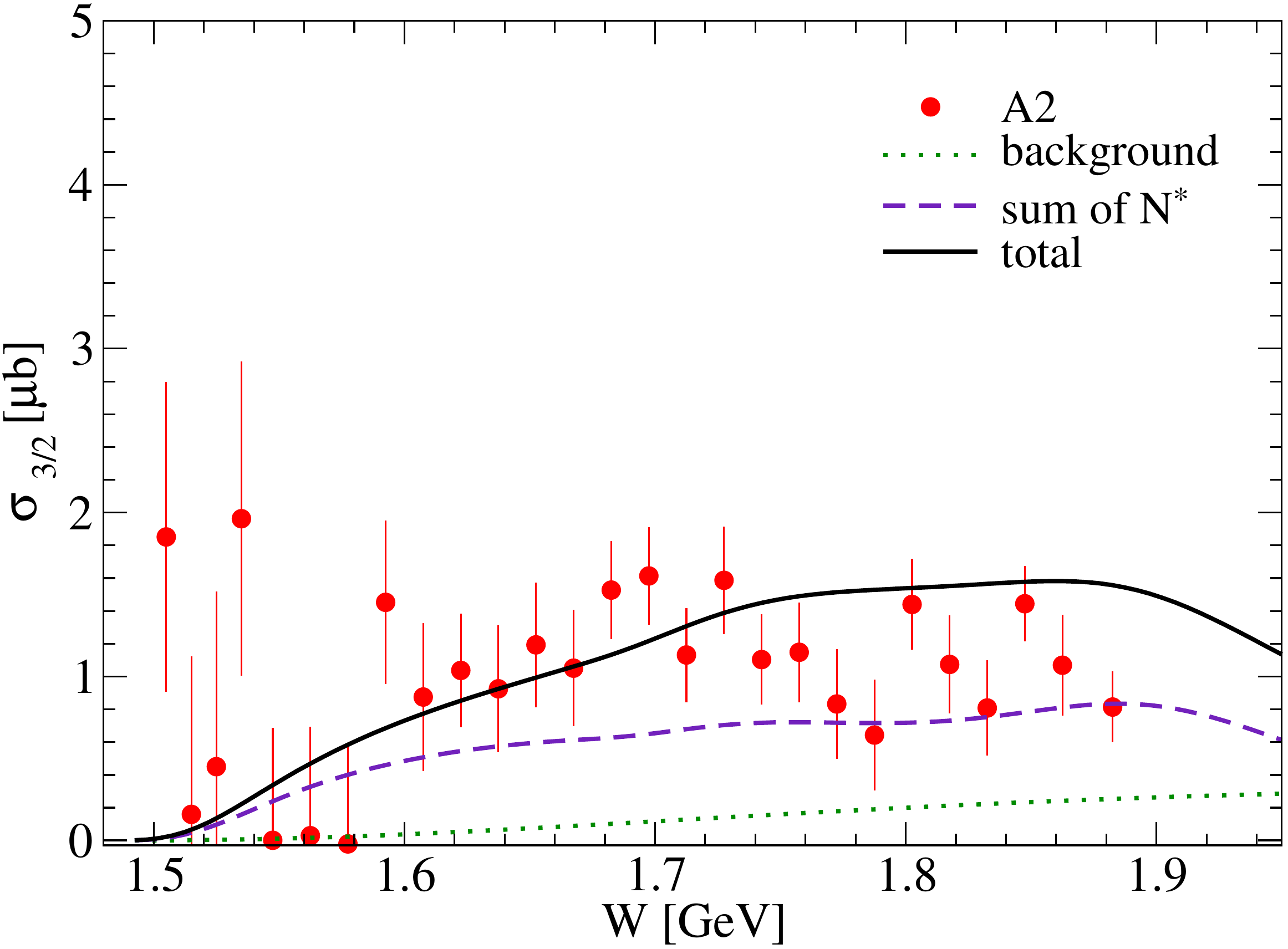}
\caption{Helicity-dependent cross sections for $\gamma n \to \eta n$.
  The circles denote the A2 data~\cite{Witthauer:2017get}.}
\label{fig:3}
\end{figure}

To clarify the role of spin $J=1/2$ and $J \geq 3/2$ $N^*$ resonances
separately, we draw the helicity-dependent cross sections in Fig.~\ref{fig:3}.
Both the $\sigma_{1/2}$ and $\sigma_{3/2}$ are in good agreement
with the A2 data~\cite{Witthauer:2017get}.
It turns out from the $\sigma_{1/2}$ that the background term and
$N(1535,1/2^-)$ interfere constructively but destructively with 
$N(1685,1/2^+)$ and $N(1710,1/2^+)$. The $N^*$ resonances with higher
spin $J \geq 3/2$ also have certain effects as found from the
$\sigma_{3/2}$. The constructive interference effect between the
backgound and higher-spin $N^*$ contributions plays an important
role in explaining the data. It implies that the $N^*$ resonances with
spin $J=1/2$ and $J \geq 3/2$ are all correctly treated in the present
model.  That is, the $N^*$ resonances with spin $J=3/2$, i.e.,
$N(1520,3/2^-)$, $N(1720,3/2^+)$, and $N(1900,3/2^+)$ give important
contributions to the $\gamma n \to \eta n$ reaction besides the
dominant spin-1/2 $N^*$ resonances $N(1535,1/2^-)$, $N(1685,1/2^+)$,
and  $N(1710,1/2^+)$. 

\section{Conclusion and outlook}
In this talk, we presented a recent investigation of $\eta n$
photoproduction, employing an effective Lagrangian approach combining 
with a Regge model. The total and helicity-dependent cross sections
were discussed in comparison with the A2 data. It was found that the
$N(1685,1/2^+)$ plays a crucial role for the description of the
$\gamma n \to \eta n$ reaction. This present results favor the
existence of the narrow $N(1685,1/2^+)$ together with our recent work
on the $\gamma n \to K^0 \Lambda$ reaction where a clue on the
evidence of the $N(1685,1/2^+)$ was also given~\cite{Kim:2018qfu}.  

Finally, we want to mention that a detailed analysis on the helicity
amplitudes in the partial-wave expansion is required in order to
elucidate the existence of the narrow $N^*(1685)$ resonance. The
corresponding work is under way. 
\vspace{0.5cm}

H.-Ch.K is grateful to M. V. Polyakov for critical comments and
criticism. This work was supported by the National Research Foundation
of Korea (NRF) grant funded by the Korea government(MSIT)
(No. 2018R1A5A1025563).

%=========================================================================
\end{document}